\documentclass[aps,pre,showpacs,superscriptaddress,twocolumn]{revtex4}
\usepackage{amsfonts}
\usepackage[dvips]{graphicx}

\usepackage{hyperref} 
\usepackage{epsfig}
\usepackage{color}
\usepackage{mathrsfs} 

\usepackage{syntonly}
\usepackage{float} 
\usepackage{amsthm,graphics,amssymb,amsmath,latexsym}
\usepackage{fancyhdr}
\begin{document}
\title{Bouncing solutions from generalized EoS}
\author{ F. Contreras}

\affiliation{Departamento de Matem\'aticas, Universidad de Santiago de Chile, Casilla 307, Santiago, Chile}

\author{ N. Cruz}
\email[e-mail: ]{norman.cruz@usach.cl}  \author{ G. Palma}

\affiliation{Departamento de F\'{\i}sica, Universidad de Santiago, Casilla 307, Santiago, Chile}


\begin{abstract} 
We present an exact analytical bouncing solution for a closed universe 
filled with only one exotic fluid with negative
pressure, obeying a Generalized Equations of State (GEoS) of the form
$P(\rho)=A\rho+B\rho^{\lambda}$, where $A$, $B$ and $\lambda$ are
constants.  In our solution $A=-1/3$ and $\lambda=1/2$ and $B<0$ 
is kept as a free parameter. For particular values of the initial
conditions, we obtain that our solution obeys  
Null Energy Condition (NEC), which allows us to
reinterpret the matter source as that of a real scalar field,
$\phi$, with a positive kinetic energy and a potential $V(\phi)$. We
compute numerically the scalar field as a function of time as well
as its potential $V(\phi)$, and find an analytical function for the potential
that fits very accurately with the numerical results obtained. The
shape of this potential can be well described by a Gaussian-type of
function, and hence, there is no spontaneous symmetry minimum of $V(\phi)$.  
We further show that the bouncing scenario is structurally stable
under small variations of the parameter $A$, such that a family of bouncing
solutions can be find numerically, in a small vicinity of the value $A=-1/3$.
\end{abstract}

\maketitle

\section{Introduction}

\hspace{-10.5pt}Non singular cosmologies such as the described by an emergent or
bouncing universe have been studied during the last decades as
alternative scenarios to the inflationary paradigm, which is the
most accepted one to describe the early universe \cite{Lemoine},
\cite{Martin}. Nevertheless, in inflation the problem of the initial
singularity still remains \cite{Borde}.  On the other hand, the
scale-invariant spectrum of cosmological perturbations can be
obtained in most inflationary models and one natural question is if
in these non singular scenarios an scale-invariant spectrum can also
be obtained.

In the case of \textit{bouncing} models the universe has emerged
from a cosmological bounce, where the scale factor takes a non-zero
minimum value so there is no initial singularity. Bouncing universes
have been investigated in a wide variety of frameworks, which
includes among others, higher order theories of gravity,
scalar-tensor theories and braneworlds. See \cite{Novello} for a
detailed discussions of different approaches to obtain bouncing
solutions.

In this paper our aim was to find bouncing solutions for a universe
only filled with one exotic fluid with negative pressure, obeying a
GEoS. A wide variety of
cosmological models have been investigated considering a GEoS of the
form
\begin{equation}
P(\rho)=A\rho+B\rho^{\lambda}, \label{EoS1}
\end{equation}
where $A$, $B$ and $\lambda$ are constants. In the framework of
general relativity the inclusion of Eq.(\ref{EoS1}) has been used to
describe the behavior of the cosmic fluid components at early and
late times, as well as the possible present phantom epoch. For
example, at early times and aiming to extend the range of known
inflationary behaviors, Barrow ~\cite{Barrow} assumed a GEoS with
$A=-1$ and $B>0$, which corresponds to the standard EoS of a perfect
fluid $p=(B -1)\rho$ when $\lambda =1$. A non singular flat universe
was found for the case $\lambda =1/2$ and $B <0$, representing an
emergent cosmological solution. It is interesting to mention that
the doubled exponential behavior of this solution was previously
found for a bulk viscous source in the presence of an effective
cosmological constant~\cite{Barrow1}. This is a consequence of the
inclusion of bulk viscosity in the Eckart´s theory, which leads to a
viscous pressure $\Pi$ of the form $-3\xi H$, where $\xi$ is assumed
usually in the form $\xi=\xi_{0} \rho^{\lambda}$. Other emergent
flat solutions were found by Mukherjee \textit{et
al}~\cite{Mukherjee} for $A >-1$ and $B
>0$.

The GEoS represented in Eq.(\ref{EoS1}) can also be seen as the sum
of the standard linear EoS $p=A\rho$ and a polytropic EoS with the
polytropic exponent $\lambda= n/(n+1)$, where $n$ is the polytropic
index. Non singular inflationary scenarios were investigated
in~\cite{Chavanis} taken particular values for $A$, $B$, and $n>0$.

In the study of late time evolution of the universe, it has been
also assumed GEoS of the type given by Eq.(\ref{EoS1}), motivated by
the fact that the constraints from the observational data implies
$\omega \approx -1$ for the EoS of the dark energy component, if it
is ruled by a barotropic EoS. Nevertheless, the values $\omega <
-1$, corresponding to a phantom fluid, or $\omega > -1$,
corresponding to quintessence can not be discarded. Within a
phenomenological approach to phantom fluids, a GEoS of the form
$p=-\rho -f(\rho)$, with $f(\rho)>0$, was proposed
in~\cite{Odintsov}. To overcome the hydrodynamic instability of a
fluid with an EoS $p=w\rho$, with $w=const<0$, a general linear EoS
of the form $p=A(\rho-\rho_{0})$ was postulated in~\cite{Babichev},
being $A$ and $\rho_{0}$ constant and free parameters. This EoS
corresponds to the particular choice $\lambda =0$ and $B=A\rho_{0}$
and was investigated as a dark fluid filling the universe.  A
bouncing solution was obtained when $1+A <0$ and $A\rho_{0}<0$. For
a Bianchi-I cosmology, the inclusion of a perfect fluid obeying a
GEoS with $\lambda=2$ leads to a great suppression on the
anisotropies in the contracting phase of a bouncing
cosmology~\cite{Bozza}.

The case with $A=-1$ and $\lambda=1/2$ was considered in
~\cite{Odintsov} and~\cite{Stefancic}. In both works the
cosmological solutions of dark energy models with this fluid was
analyzed, focusing in the future expansion of the universe. A late
time behavior of a universe filled with a dark energy component with
an EoS given by Eq.(\ref{EoS1}) has been investigated in
~\cite{Paul}, ~\cite{Paul1}, where the allowed values of the
parameters $A$ and $\gamma$ were constrained using H(z)-z data, a
model independent BAO peak parameter and cosmic parameter (WMAP7
data).

Also theoretical studies like the so called running vacuum energy in
QFT (see ~\cite{Shapiro}) gives rise to a cosmological constant with
a dynamical evolution during the cosmic time, which allows to
conclude that GEoS of the type of Eq.(\ref{EoS1}) could also
effectively represent these scenarios under some specific
assumptions.

In this work we use a rather conservative setup introducing a
positive curvature and the particular values $A=-1/3$ and
$\lambda=1/2$, letting $B<0$ as a free parameter of the model. With
this election the strong energy condition is violated, which is a
condition to have bouncing solutions, but NEC
holds, and thus our particular GEoS has a parameter $\omega$ 
that evolves with the cosmic time, but lies in the range of 
quintessence fluids, for some choice of the initial conditions,
except for $t\rightarrow \pm \infty$, where the fluid behaves like a
cosmlogical constant. These particular values of $A$ and $\lambda$
allow to find an exact analytical bouncing solution for the scale
factor.

Reinterpreting the matter source in terms of a real scalar field, we
can compute numerically the scalar field and its potential. We also
found an analytical expression for this potential that fit very
accurately the numerical solutions, with a coefficient of determination ($r^{2}$) equal to $r^{2}=0.99999$.

We also study the robustness of the bouncing solution when the GEoS
is modified by including a perturbative term in the standard linear
coefficient $A$. We find that under reasonable constraints on the
perturbative parameter, the solution is analytic in $\epsilon$ and
the first order correction allows to extend the behavior of the
bouncing solution beyond the value $-1/3$, for which an explicit
analytic solution was found. The perturbative expansion leads as
well to conclude that the properties of the scalar potential (shape
and minimum) proposed as source for the effective equation of state
are stable, provided $\epsilon$ remains small enough.

This paper is organized as follows. In section II we present the
particular considered GEoS and show the analytical bouncing solution
found and their main properties. In particular, we present the
evolution of the parameter $\omega$ with the cosmic time, discussing
its quintessential behavior and how this allow to describe the
matter content by a usual real scalar field with a potential. In
section III we evaluate numerically the scalar field and its
potential associated to our exact solution. We also find an
analytical expression for this potential that fits the numerical
results found. In section IV we investigate the stability of the
bouncing solution when the GEoS is modified disturbing the parameter
$A$ by an small quantity. So in this case, we make a study of
structural stability under small variations of the parameter $A$.
Finally, in section V we discuss some features and their further possible
applications to suitable bouncing models.

\section{Exact bouncing solution from GEoS}

\hspace{-10.5pt}In what follows we will discuss an analytical solution for a closed
universe found in~\cite{Contreras}, for the case in which the
parameter $A$ takes the value $-1/3$ and $\lambda=1/2$ in Eq.(\ref{EoS1}). As we
discuss bellow the $\omega = p/\rho$ parameter can represent phantom
and quintessence fluids, depending on the initial conditions. This
exact solution describe a bouncing universe, assuming that one
fluids with a GEoS is present in the early universe.

For a universe with positive curvature ($k=1$), the equation of
constraint of the Friedmann equations is given by
\begin{equation}
\rho=3\left( \frac{\dot{a}}{a}\right) ^{2}+\frac{3}{a^{2}},
\label{constraint}
\end{equation}
and the equation of continuity by
\begin{equation}
\dot{\rho}+3H( \rho +p) =0. \label{continuity}
\end{equation}

Using the change of variable $s=-\dfrac{2}{B\sqrt{3}}$  the GEoS of Eq. (\ref{EoS1}) can 
be rewritten as	
\begin{equation}
p(\rho)=-\frac{1}{3}\rho-\dfrac{2}{s\sqrt{3}}\rho^{1/2}.
\label{EoS2}
\end{equation}

Solving the above equations one finds the following solution
\begin{equation}
a(t)=s\left[ \cosh\left(
\frac{t-t_{0}}{s}\right)-c\right], \label{exactsolution}
\end{equation}
where $t_{0}$ and $c$ are integration constants. The
bouncing solution is obtained when $s>0$. This solution
represents a universe expanding exponentially for $t \in
(-\infty,\infty)$. The scale factor takes a minimum value
$a(t=t_{0})= s(1-c)$. The positivity of
the scale factor constraints $c$ to be in the following range
$(-\infty ,1)$. Before to express $c$ in
terms of the initial energy density we evaluate $H$, $\dot{H}$ and
$\ddot{H}$ using the Eq.(\ref{exactsolution}). Their expressions are
the following:

\begin{gather}
H(u)=\frac{\sinh(u)}{s(\cosh(u)-c)}, \label{Hdet} \vspace{2mm} \\
\dot{H}(u)=\frac{1-c \cosh{ \left(u\right)}}{ \left[ s(\cosh { \left(u\right)}-c) \right]^{2}}, \vspace{2mm} \\
\ddot{H}(u)=-\frac{\sinh{(u)}\left( 2-c^2\right)+\sinh{(2u)\left( \dfrac{r}{3}+\dfrac{1}{s}\right)}}{\left[ s(\cosh{(u)-c)}\right]^{3}},
\end{gather}
where 

\begin{equation}
u={t-t_{0}\over s}.
\label{u(t)}
\end{equation}

For $c <0$ the Hubble parameter
is a strictly increasing function, so there are no critical points and we have
 $ H(t\rightarrow -\infty)= -\frac{1}{s}$ and $ H(t\rightarrow \infty)= \frac{1}{s}$, 
then for late times this solution behaves like a de Sitter universe.

It is straightforward to evaluate the energy density as a function
of the cosmic time using Eq.(\ref{constraint}) and the expression
for $a(t)$ and $H(t)$ given by Eq.(\ref{exactsolution}) and
Eq.(\ref{Hdet}), respectively. The expression for the energy density
is then given by
\begin{equation}
\rho=\frac{3\cosh^{2}(u)}{\left[ s(\cosh(u)
-c)\right]^{2}}. \label{rhodet}
\end{equation}

Using the initial conditions $a(t_{0})=a_{0}$ in (\ref{exactsolution}) and $\rho (t_{0})=\rho _{0}$ in (\ref{rhodet}) we obtain that

\begin{equation}
\begin{array}{c}
a_{0}=s(1-c), \quad \ \ 
3=a_{0}^{2}\rho _{0}. 
\end{array}
\label{condinit}
\end{equation}

One dimensional restoration lead to the Eq.(\ref{condinit}) takes the following form:

\begin{equation}
\begin{array}{c}
a_{0}=s(1-c)\dfrac{v^{2}}{R_{0}\sqrt{8\pi G}}, \quad \ \ 
\dfrac{v^{2}}{R_{0}^{2}}=\dfrac{8\pi G}{3}\rho _{0}a_{0}^{2},
\end{array}
\end{equation}
where $v$ is the speed of light, $R_{0}$ is the radius of curvature, $G$ is the gravitational contant and $\rho$ is the energy density. Because the model considers the universe with  curvature positive, the radius the curvature $R_{0}$ also is a free parameter.  

A very special situation occurs for $c =0$ or equivalently
$\rho_{0}=3/s^{2}$ and $a_{0}=s$, because the energy density preserves this
constant value during all the cosmic evolution. It means that for a
closed universe with the EoS that we are considering, the universe
expand with acceleration but the energy density remains constant,
like in de Sitter solution for a closed universe. Note that the
Hubble parameter for the de Sitter solution is given by
\begin{equation}
H(t)=\sqrt{\frac{\lambda}{3}}\tanh \left ( \sqrt{\frac{\lambda
}{3}}(t-t_{0})\right )   , \label{HdeSitter}
\end{equation}
and in our case the Eq.(\ref{Hdet}) with $c =0$ takes the similar
form:
\begin{equation}
H(t)=\frac{1}{s}\tanh \left (\frac{t-t_{0}}{s} \right ).
\label{HsimilardeSitter}
\end{equation}

In the next subsection we give an interpretation of the used GEoS in terms of well known fluids.

\subsection{\bf{Fluid sources of the bouncing solution}}

We can obtain the energy density as a function of the scale factor
if we replace the Eq.(\ref{exactsolution}) in the Eq.(\ref{rhodet})
\begin{equation}
\rho(a)=3\left( \frac{1}{s}+\frac{c}{a}\right)^{2}.
\label{rhodeaEoS2}
\end{equation}

Introducing Eq.(\ref{rhodeaEoS2}) in Eq.(\ref{EoS2}) we obtain the
fluid pressure as a function of the scale factor
\begin{equation}
p(a)=-\left( \frac{1}{s}+\frac{c}{a}\right)^{2}-\dfrac{2}{s}\left( \frac{1}{s}+\frac{c}{a}\right).
\label{pdeaEoS2}
\end{equation}

Expanding the terms of the above both expressions yields:

\begin{gather}
\rho(a)=\frac{3}{s^2}+\frac{6c}{sa}+\frac{3c^{2}}{a^{2}}=\rho_1+\rho_2+\rho_3, \label{rhoaextend} \vspace{2mm} \\
p(a)=-\frac{3}{s^2}-\frac{4c}{sa}-\frac{c^{2}}{a^{2}}=p_1+p_2+p_3.
\label{Paextend}
\end{gather}

Comparing each terms of the expansions our fluid can be seen as the
sum of three fluids with the EoS given by $\omega_{1}=
p_{1}/\rho_{1}=-1$, $\omega_{2}= p_{2}/\rho_{2}=-2/3$ and
$\omega_{3}= p_{3}/\rho_{3}=-1/3$, respectively.  So the first fluid
corresponds to a cosmological constant, the second one is a quintessence
and the last corresponds to a fluid which drives an expanding universe with zero acceleration. Notice that in the above descomposition the EoS of each fluid is constant. Therefore, each $\omega _{i}$ is independent of the parameters $s$, $t_{0}$ and $c$.

Lets us evalue the EoS, $\omega = p/\rho$, for this fluid, which in
terms of the cosmic time takes the expression
\begin{equation}
\omega=-1+\dfrac{2c}{3\cosh \left({t-t_{0}\over s}\right) }. \label{omegabouncing}
\end{equation}

\begin{figure}[H]
\begin{center}
\includegraphics[scale=0.5]{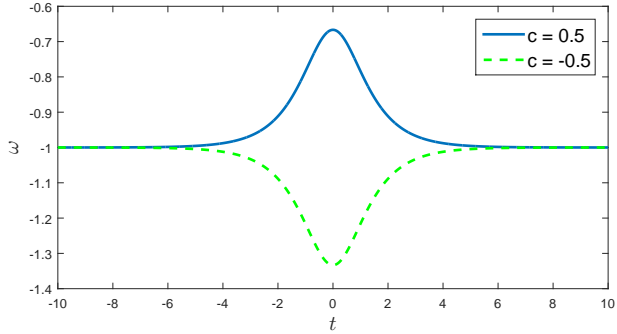}
\caption{Plot of the parameter  $\omega $  given by Eq.(\ref{omegabouncing}), for the parameter values $c=0$.$5$, $s=1$ and $t_{0} =0$.}
\label{Fig:parestado}
\end{center}
\end{figure}

The lower plot in Fig. \ref{Fig:parestado}.
depicted the behavior of this parameter $\omega $.

Note that the EoS becomes like a cosmological constant $\omega \rightarrow -1$ for $t\rightarrow \pm \infty$. For the case with $c =0.5$
the fluid ruled by the EoS given in Eq.(\ref{omegabouncing}) behaves
like quintessence for the lapse associated
at the time of bouncing. For the case $c=-0.5$ the EoS behaves like a phantom fluid within the period associated to the bouncing. We will focus on the particular interval $1>c>0$  because in this case the GEoS leads to a quintessence-type of behavior. With $c$ within this range, the matter content of the universe  
can be described by a real scalar field with a lagrangian minimally coupled to gravity given by
\begin{equation}
\mathcal{L}= \frac{1}{2} \phi_{,\mu}\phi^{,\mu} -V(\phi).
\label{Lagrang}
\end{equation}

In order to reinterpret the matter source as that of a
scalar field, we will evaluate the scalar field $\phi$ and the potential $V(\phi )$ in the next section.

\section{Computation of the scalar field and its potential}

\hspace{-10.5pt}If we consider the matter content of the universe modeled by a perfect fluid, then the density and pressure in term of the scalar field are given by

\begin{equation}
\begin{array}{c}
\rho = \dfrac{1}{2}\dot{\phi} ^{2}+V(\phi ),\quad
p=\dfrac{1}{2}\dot{\phi} ^{2}-V(\phi ).\\ \\
\end{array}
\label{eccamppot}
\end{equation}

Using the Eq.(\ref{EoS2}) and the Eq.(\ref{rhodet}) in the Eq.(\ref{eccamppot}) we obtain:

\begin{gather}
\dfrac{d\phi}{du}=\pm \dfrac{\sqrt{2c\cosh (u)}}{\cosh (u)-c},
\label{dphidu} \vspace{2mm} \\
V(u)=\dfrac{\cosh (u)(2\cosh (u)-c)}{\left[ s(\cosh (u)-c)\right] ^{2}},
\label{Vu}
\end{gather}
where  $u$ is given by the Eq.(\ref{u(t)}). 
Integrating the Eq.(\ref{dphidu}), the function $\phi $ is obtained as:

\begin{equation}
\phi(\text{u})=-i\sqrt{8c}\left[ F\left( \dfrac{i\text{u}}{2},\sqrt{2}\right) +\dfrac{c}{1-c}\Pi \left( \dfrac{i\text{u}}{2}, \dfrac{2}{1-c}, \sqrt{2}  \right) \right],
\label{phiuintelip}
\end{equation}
where $i$ is the imaginary unit, $F$ is the elliptic integral of the first kind and $\Pi$ is the elliptic integral of the third kind. Both are defined in \cite{Gradshteyn}.

The function $d\phi \over du$ in Eq.(\ref{dphidu}) is a continuous function. Therefore, this function have a real primitive function, but this can't be represented by  elementary functions. Thus the imaginary value in $\phi (u)$ in the Eq.(\ref{phiuintelip}) is only a artifice of the representation of the function. 

\begin{figure}[H]
\begin{center}
\includegraphics[scale=0.5]{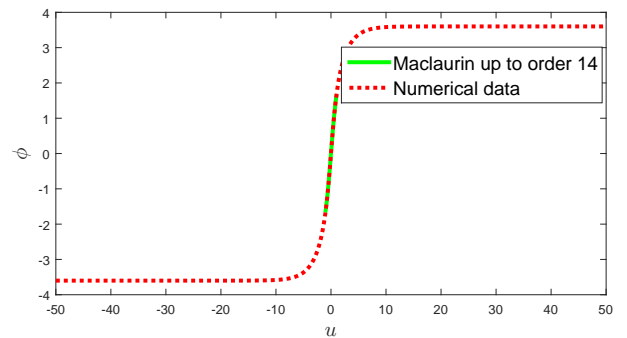}
\caption{Plot of $\phi (u)$ using the Maclaurin series obtained from Eq.(\ref{phiuserie}) and its numerical solution for the parameter values $c=0$.$5$ and $s =1$.}
\label{fig:discrevsseriephi0c05}
\end{center}
\end{figure}

In order to numerically obtain $\phi(u)$ from the above equations, we have used standard integration subroutines from Matlab, imposing for consistence the initial condition $\phi(0)=0$. The result is displayed in Fig. \ref{fig:discrevsseriephi0c05}, where for comparison we have also plotted its Maclaurin expansion up to order 14th, whose coefficients are obtained in Appendix A. In addition, in this appendix the convergence radius of this series shown to be $\arccos (c)$. A remarkable agreement among both results is found, within the common range of $u$, which represents a severe test of accuracy to the numerical solution.

Moreover, we have compared the results for $\phi(u)$ obtained by the Maclaurin series with the one obtained by the numerical integration by computing the Pearson's coefficient $r=0.999985$, which allows to use the numerical solution beyond the convergence radius of the series expansion.

\begin{figure}[H]
\begin{center}
\includegraphics[scale=0.5]{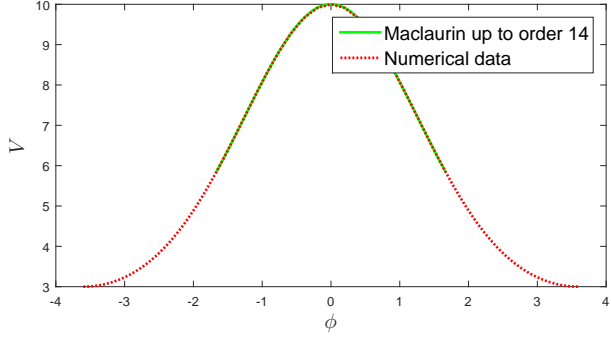}
\caption{Plot of $V(\phi)$ using the Maclaurin series obtained in the Appendix A Eq.(\ref{coefbetasubkgenbounc}), as well as by the numerical integration from Eq.(\ref{Vu}), for the particular parameter values $c=0$.$5$ and $s =1$.}
\label{fig:discrevsserieV0c05}
\end{center}
\end{figure}

We have also computed numerically $V(\phi )$ from Eq.(\ref{Vu}) as well as its Maclaurin expansion using the expression deduced in Appendix A (see the Eq.(\ref{coefbetasubkgenbounc})). Similarly to the analysis explained above, we have measured the degree of agreement among both methods by computing the Pearson's coefficient, which in this case is 1 ($r=1$). Both results are displayed in Fig. \ref{fig:discrevsserieV0c05}.

In the next subsecction we will find analytical expressions for the field $\phi $ and its potencial $V$ by performing high accuracy fits.

\subsection{\bf{ Analytical representation of $\phi(u)$} and $V(\phi)$}

\begin{figure}[H]
\begin{center}
\includegraphics[scale=0.5]{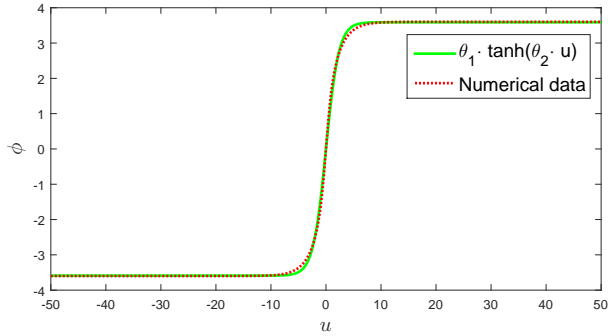}
\caption{Plot of $\phi (u)$ obtained from its numerical solution as well as from the fit given by Eq.(\ref{fitfieldtanh}) for the values $c=0$.$5$ and $s =1$. }
\label{fig:compajustecampoc05}
\end{center}
\end{figure}

In order to characterize analytically the shape of the scalar potential, and eventually to compare it with other quintaessence potentials, we perform a fit of the numerical data for $\phi(u)$ using the function $\tanh(x) $, 

\begin{equation}
\phi (u)= \theta_{1}\cdot \tanh (\theta_{2}\cdot u),
\label{fitfieldtanh}
\end{equation}

where $\theta_{1}$ and $\theta_{2}$ are paremeters positives of fit. The fit of field $\phi$ can be observed in the Fig. \ref{fig:compajustecampoc05}.

The fit represented by Eq.(\ref{fitfieldtanh}) is of high quality as the corresponding 
coefficient of determination is $r^{2}=0.99981$, for the fit parameters $\theta_{1}\approx 3.5892$ 
and $\theta _{2}\approx 0.42631$.

We have also fitted the numerical data for the field $\phi (u)$ to the function $\theta_{1} \arctan (\theta_{2} u )$, where $\theta _{1}$ and $\theta_{2}$ are the fit parameters, but the quality of the fit was not quite comparable to the one obtained by using the analytic form given by Eq.(\ref{fitfieldtanh}). 

\begin{figure}[H]
\begin{center}
\includegraphics[scale=0.5]{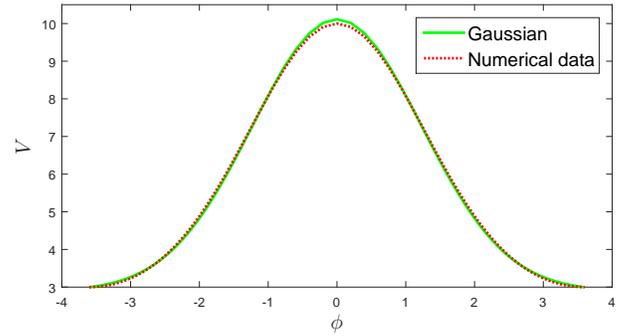}
\caption{Plot of the numerical data for $V (\phi )$ and the Gaussian fit defined by Eq.(\ref{fitVu}), for the values $c=0$.$5$ and $s =1$.}
\label{fig:AjusteVgaussV3parc05}
\end{center}
\end{figure}

Because of the shape of the scalar potential $V(\phi )$, we have used a Gaussian as a trial function :
\begin{equation}
V(\phi)=\sigma _{1}\cdot \exp \left(-\sigma_{2}\cdot \phi ^{2} \right)+\sigma_{3},
\label{fitVu}
\end{equation}
where $\sigma_{1}$, $\sigma_{2}$ and $\sigma_{3}$ are fit parameters. The result of this fit is displayed in Fig. \ref{fig:AjusteVgaussV3parc05}.

As it is shown in Fig. \ref{fig:AjusteVgaussV3parc05}, the Gaussian function fits very accurately the numerical data, moreover the coefficient of determination $r^{2}=0.99965$ for the fit parameters  $\sigma _{1}\approx 7.2164$, $\sigma _{2}\approx 0.32953$ and $\sigma _{3}\approx 2.896$. 

Due to the bounded domain of the potential $ V (\phi) $, one has to modify the Gaussian such that it falls off to zero as the field $\phi$ approaches the limits $ \pm \phi _{ max}$. We have fulfilled this constraint by introducing the modified trial function:

\begin{equation}
V(\phi )=(V_{max}-V_{min}) \exp \left[ \displaystyle -\sigma_{1} \phi ^{2}-\dfrac{\sigma_{2} \phi^{2}}{\phi_{max}^{2}-\phi ^{2}}\right] +V_{min}
\label{Ajustegausscut}
\end{equation}
where $\sigma_{1}$ and $\sigma_{2}$ are the fit parameters. $ \phi_{max}$ turns out to be $\phi _{max}\approx 3.6009$.  $V_{max}$ and $V_{min}$ are the maximum and minimum values of the potential $V(\phi )$. They can explicitly be obtained as follows: inserting Eqs. (\ref{rhodeaEoS2}) and (\ref{pdeaEoS2}) into Eq.(\ref{eccamppot}) one obtains: 

\begin{equation}
\begin{array}{l}
\dot{\phi }(a)=\dfrac{2c}{sa}+\dfrac{2c^{2}}{a^{2}},\quad \ \
V(a)=\dfrac{3}{s^{2}}+\dfrac{5c}{sa}+\dfrac{2c^{2}}{a^{2}}.
\end{array}
\label{sisguillermo}
\end{equation}

Since the constants $c$ and $s$ are positive, the maximum (minimum) of $V$ is obtained when $a$ is a minimum (maximum). But from Eq.(\ref{exactsolution}) the maximum and minimum values of $a(t)$ are $\infty$ and $s(1-c)$ respectively. Inserting these values into Eq.(\ref{sisguillermo}) we obtain:

\begin{equation}
\begin{array}{l}
V_{max}=\dfrac{ 3-c}{s^{2}(1-c)^2},\quad \quad 
V_{min}=\dfrac{3}{s^{2}}.
\end{array}
\label{VmaxVmin}
\end{equation}

\begin{figure}[H]
\includegraphics[scale=0.50]{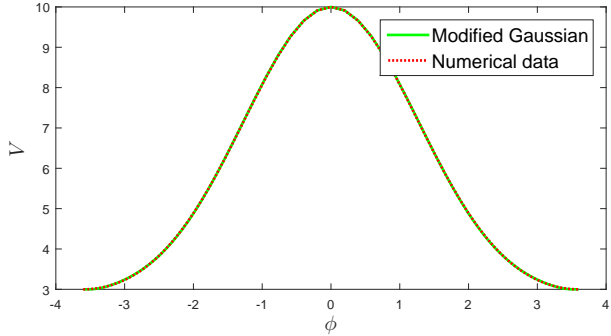}
\caption{Plot of $V (\phi )$ obtained from the numerical solution and by the fit defined by Eq.(\ref{Ajustegausscut}), for the values $c=0$.$5$ and $s =1$.}
\label{Fig:AjusteVgausscutVc0.5}
\end{figure}

The result of the fit of $V(\phi)$ using the function defined in  Eq.(\ref{Ajustegausscut}) is shown  in Fig.  \ref{Fig:AjusteVgausscutVc0.5}.

The modified Gaussian function fits remarkable well the numerical data for $V(\phi)$ as the coefficient of determination is $r^{2}=0.99999$ for the values $\sigma _{1}=0.29243$ and $\sigma _{2}=0.32674$. Considering the constraint on the domain of $\phi$ and the higher accuracy of this modified Gaussian function, we conclude that the expression given by Eq.(\ref{Ajustegausscut}) is the faithfuliest representation of $V(\phi)$.
 
\begin{center}
\subsection{\bf {Analysis of the modified Gaussian potential}}
\end{center}

\begin{figure}[H]
\includegraphics[scale=0.5]{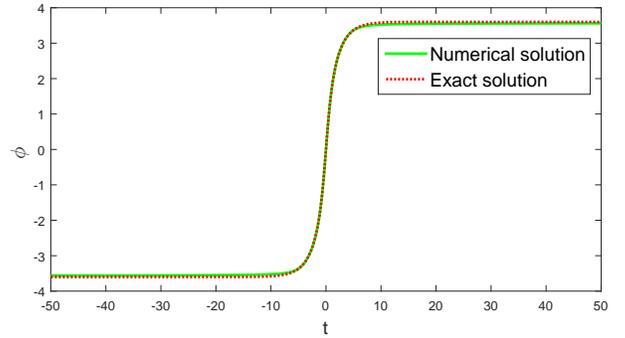}
\caption{Plot of $\phi$ as a function of time obtained numerically from Eq.(\ref{sisguillermo}) using the fitted expression for $V$ given by Eq.(\ref{Ajustegausscut}), as well as from the exact solution of Eq.(\ref{exactsolution}), for the parameter values $c=0$.$5$ and $s=1$. }
\label{Fig:AjusteVgausscutphic0.5}
\end{figure}

\begin{figure}[H]
\includegraphics[scale=0.5]{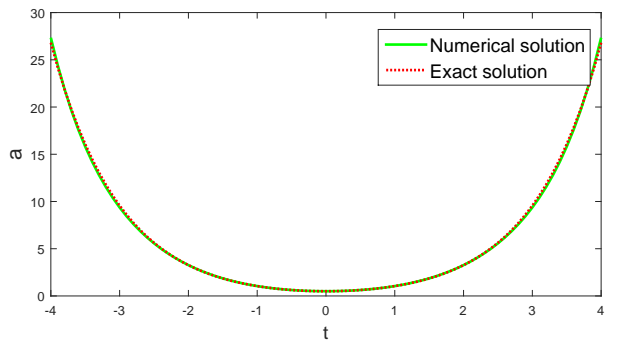}
\caption{Plot of $a(t)$ obtained of Eq.(\ref{sisguillermo}) used $V$ given by Eq.(\ref{Ajustegausscut}), as well as from the exact solution of Eq.(\ref{exactsolution}), for the parameter values $c=0$.$5$ and $s=1$. }
\label{Fig:AjusteVgaussacutac0.5}
\end{figure}

\hspace{-10.5pt}As a consistency check it is possible to start with the expression of Eq.(\ref{Ajustegausscut}) for the potential $V(\phi )$ and solve numerically the set of equations (\ref{sisguillermo}). By using standard integration subroutines this numerical strategy allows to obtain the functions $\phi (t)$ and $a(t)$, whose results are displayed in Figs. \ref{Fig:AjusteVgausscutphic0.5} and \ref{Fig:AjusteVgaussacutac0.5}, where for comparison, we have included the exact solutions given by Eqs. (\ref{phiuintelip}) and (\ref{exactsolution}) respectively.

Both figures show a remarkable agreement among the numerical solutions and the exact expressions, which is quantified by the coefficients of determination $r^{2}=0.99999$ for the scalar field $\phi(t)$, and $r^{2}=0.99962$ for the scale factor $a(t)$.

\section{Structural stability of the GEoS and prevalence of the  bouncing solution }

Since and exact bouncing solution is obtained for the very particular value $A=-1/3$, it is an open question whether variations of the parameter $A$ leads to structural stability of the field equations, or in other words whether the system still have bouncing solutions when the GEoS is modified by disturbing the parameter $A$ by $A\rightarrow A+\epsilon$ in the Eq.(\ref{EoS1}). We will explore this issue taking into account a GEoS of the form
\begin{equation}
p=\left( -{1\over 3}+\epsilon \right)\rho - \dfrac{2}{s\sqrt{3}}\rho ^{1/2},
\label{ecestadogen}
\end{equation}
where $\epsilon \ll |1/3|$. We study first perturbations on the exact solution found, driven by the initial GEoS of Eq.(\ref{EoS2}). We shall consider the following perturbations on the parameters $a$,
$\rho$ and $P$
\begin{equation}
\begin{array}{l}
a(u)=a_{(0)}(u)+\epsilon a_{(1)}(u)+O(\epsilon ^{2}),\vspace{2mm} \\\rho(u)=\rho_{(0)}(u)+\epsilon \rho_{(1)}(u)+O(\epsilon ^{2}),\vspace{2mm} \\
p(u)=p_{(0)}(u)+\epsilon p_{(1)}(u)+O(\epsilon ^{2}),
\end{array}
\label{sistgen}
\end{equation}
where for the case $\epsilon =0$ the solutions are given by the Eq.(\ref{exactsolution}) in the following form

\begin{equation}
\begin{array}{l}
a_{(0)}(u)=s\left( \cosh (u)-c\right),\vspace{2mm} \\
\rho _{(0)}(u)=\dfrac{3(\cosh (u))^{2}}{s^{2}(\cosh (u)-c)^{2}}, \vspace{2mm} \\
p_{(0)}(u)=-\dfrac{1}{3}\rho _{(0)}-\dfrac{2}{s\sqrt{3}}\rho _{(0)}^{1/2}.
\end{array}
\label{solord0}
\end{equation}

Then using Eq.(\ref{constraint}) and the continuity equation
\begin{equation}
\dot{\rho }=-3\dfrac{\dot{a}}{a}(\rho + P), \label{ecconsgen}
\end{equation}
we can obtain a system to first order in
$\epsilon$, which allows to find $a_{(1)}$ and $\rho _{(1)}$. The
system is:
\begin{equation}
\begin{array}{c}
a'_{1}(u)=A(u)a_{1}(u)+B(u)\rho _{1}(u), \vspace{2mm} \\
\rho' _{1}(u)=C(u)a'_{1}(u)+D(u)a_{1}(u)+E(u)\rho _{1}(u)+F(u),
\end{array}
\label{Sistema}
\end{equation}
where the ``prime'' ({ \large $_{'} \ $}) denotes the derivative with respect to
$u$, and the cofficients $A(u)$, $B(u)$, $C(u)$, $D(u)$, $E(u)
$, $F(u)$ are obtain in the Appendix $B$ (see the Eqs.(\ref{coeffAandB}) and (\ref{coeffC,D,EandF})). 

Solving the system using the given initial condition, 

\begin{equation}
\begin{array}{l}
a(u=0)=a_{(0)}(0),\quad \quad
\dot{a}(u=0)=\dot{a}_{(0)}(0),
\end{array}
\label{initcondi}
\end{equation}
we obtain that the first order contribution for the scale factor, the energy density and the pressure are given by 

\begin{widetext}
  \begin{equation}
  \begin{array}{c}
a_{(1)} =\dfrac{3s}{2} \left\{ -1+  \cosh (u) +c\ln \left[ M\right] -\sinh (u) \left[   u+\dfrac{2c\arctan \left( N\right)}{\sqrt{1-c^{2}}}  \right]          \right\}, \vspace{2mm} \\
\rho_{(1)}=\dfrac{9\cosh(u)}{s^{2}(\cosh (u)-c)^{3}}\left\{ \cosh (u)-\cosh ^{2}(u)+cu\sinh(u)+\dfrac{2c^{2}\sinh (u)\arctan \left( N \right)}{\sqrt{1-c^{2}}}-c\cosh (u)\ln \left[ M \right]     \right\},  \vspace{2mm} \\
p_{(1)}=\rho _{(0)}-\dfrac{\rho _{(1)}}{3}-\dfrac{\rho _{(1)}}{s\sqrt{3}\rho _{(0)}^{1/2}},\quad
\text{where } M=\dfrac{\cosh (u)-c}{1-c} \ \ \text{and } \ N=\dfrac{(1+c)\tanh (u/2)}{\sqrt{1-c^{2}}}.
  \end{array}
  \label{solfinaleps1ord}
  \end{equation}
\end{widetext}

Now, we analyze the functions $a_{1}$ and $\rho _{1}$ of above equation. Let us consider the first two  nonzero terms of the Maclaurin expansion of Eq.(\ref{solord0}), which represents $a_{0}$, and the first nonzero nonzero of the Maclaurin expansion of the expression for $a_{1}$, which is given by Eq.(\ref{solfinaleps1ord}). We obtain that

\begin{equation}
a_{0}(u)=s(1-c)+\dfrac{su^{2}}{2},\hspace{1cm} a_{1}(u)=-\dfrac{3su^{2}}{4(1-c)}.
\end{equation}

Note that $a_{1}(0)=0$ and that the sign of $a_{1}$ is always negative contrary to $a_{0}$ that is positive. For this reason for $\epsilon >0$ the scale factor of the bouncing universe in a neighborhood of $u=0$ grows lesser than the original solution and for $\epsilon<0$ the scale factor grows fasther than the original one. This behavior can be seen from Eq.(\ref{ecestadogen}), since for $\epsilon >0$ the GEoS is the same that the original GEoS plus the term $\epsilon \rho$. Therefore, the expected behavior of the new scale factor should be less pronunced that the scale factor of the original solution, because the new quintessence fluid have one state parameter $\omega $ greater than original. Finally, if we capare magnitudes of the order $u^{2}$ for $a_{0}$ and $a_{1}$, and if in addition we include the term $\epsilon $ in $a_{1}$, we obtain that 

\begin{equation}
\left|\dfrac{s}{2}\right|>|\epsilon|\left|-\dfrac{3s}{4(1-c)}\right| \Leftrightarrow c<1-\dfrac{3|\epsilon|}{2}.
\label{relata_0a_1}
\end{equation}

Therefore, we will have a bouncing behavior for $u\approx 0$ for any couple of costants $c$ and $\epsilon $ that satisfy the Eq.(\ref{relata_0a_1}). Note that for any $c$ between $(0,1)$ always exist one $\epsilon >0$ and $|\epsilon| \ll {1\over 3}$ which satisfy the above equation.

Now, let's analyze $\rho_{1}$ of Eq.(\ref{solfinaleps1ord}). Evaluating the first nonzero terms of Maclaurin series of $\rho_{0}$ and $\rho_{1}$, we obtain

\begin{equation}
\rho_{0}= \frac{3}{s^{2}(1-c)^2}-\frac{3 c u^2}{s^{2}(1-c)^3 }, \hspace{0.2cm} \rho_{1}=-\frac{9 (1-2 c) u^2}{2 s^{2}(1-c)^4 }.
\label{relat2rhorho1}
\end{equation}

Note that $a_{1}(0)=\rho_{1}(0)=0$ and also that the sign of $\rho_{1}$ is negative for $c<0.5$ and positive for $c>0.5$. In the case $c=0.5$ the first $\rho _{1}=0$. Thus, for any $|\epsilon |\ll {1\over 3}$, $\rho_{0}$ will dominate over $\rho_{1}$ for $u\approx 0$.  Now, if we compare magnitudes of the order $u^{2}$ for $\rho_{0}$ and $\rho_{1}$, and if in addition we include the term $\epsilon $ in $\rho_{1}$, we obtain that for $c\neq 0.5$

\begin{equation}
\left|\frac{3 c }{s^{2}(1-c)^3 }\right| >|\epsilon|\left|\frac{9 (1-2 c) }{2 s^{2}(1-c)^4 }\right| \Leftrightarrow \dfrac{2c(1-c)}{3|1-2c|}>|\epsilon|.
\label{relatrho0rho1}
\end{equation}

This equation tell us what values of $c$ and $\epsilon$ lead  $\rho _{0}$ dominates over $\rho_{1}$ for $u\approx 0$. 

\begin{figure}[H]
\includegraphics[scale=0.5]{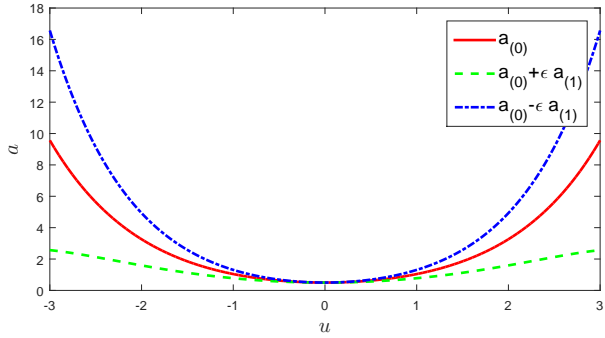}
\caption{Plot of the scale factor and its first order correction, obtained from Eq.(\ref{solfinaleps1ord}), for the values $\epsilon =0$.$15$, $c=0$.$5$ and $s=1$. }
\label{Fig:facescperc0.5}
\end{figure}

The behavior of original scale factor and the perturbed ones are shown in Fig. \ref{Fig:facescperc0.5}. The $ \epsilon $ that we consider in the graphics corresponds to a $ 45 \% $ of $ 1/3 $ and $c=0.5$. With these $c$ and $\epsilon$ we obtain that the Eq.(\ref{relata_0a_1}) is satisfied and therefore, there is a bouncing behavior. 

\begin{figure}[H]
\includegraphics[scale=0.5]{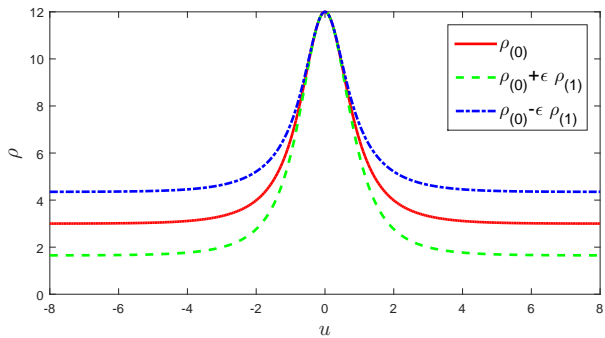}
\caption{Plot of energy density $\rho$ and its first order correction obtained from Eq.(\ref{solfinaleps1ord}), for the parameter values $\epsilon =0$.$15$, $c=0$.$5$ and $s=1$. }
\label{Fig:densidadperc0.5}
\end{figure}

The coefficient of determination is $r^{2}=0.94759$, for $a_{0}$ and $a_{0}+\epsilon a_{1}$, and $r^{2}=0.99875$, for $a_{0}$  and $a_{0}-\epsilon a_{1}$. The energy density $\rho _{1}$ given by Eq.(\ref{solfinaleps1ord}) is shown in Fig. \ref{Fig:densidadperc0.5}.

The coefficient of determination is $r^{2}=0.99980$, for $\rho_{0}$ and $\rho_{0}+\epsilon \rho_{1}$ and $r^{2}=0.98566$, for $\rho_{0}$ and $\rho_{0}-\epsilon \rho_{1}$. We can note that despite of the high value of $\epsilon $, the behavior of scale factor and of the energy density perturbed to first order are quite similar to the obtained with the exact solution.   Moreover, if we change the parameters $c$ and $s$ in their respective domain we will get universes keeping the same shape of the bouncing but with different growths. 

\begin{figure}[H]
\includegraphics[scale=0.5]{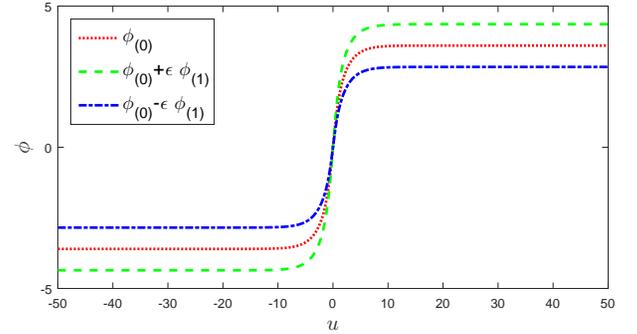}
\caption{Plot of the scalar field $\phi$ and its first order correction, for the parameter values $\epsilon =0$.$15$, $c=0$.$5$ and $s=1$. }
\label{Fig:pertepscampoc05}
\end{figure}

We evaluate numerically the first order perturbation of the field $\phi$, which we denoted by $\phi _{1}(u)$. Its behavior is showed in the Fig. \ref{Fig:pertepscampoc05}.

Like the scale factor and the energy density, we obtain that the first order perturbed field $\phi_{1} $ preserve the shape of unperturbed solution $\phi _{0}$. The coefficient of determination is $r^{2}=0.99999$, for $\phi_{0}$ and $\phi _{0}+\epsilon\phi_{1}$,  and $r^{2}=0.99998$ for $\rho _{0}$ and $\rho _{0}-\epsilon \rho_{1}$. \\ 

To obtain the potential $V_{1}$ we exapand the function $V(\phi )$ in $\epsilon $. From this expansion  we obtain the following expression

\begin{equation}
V(\phi)= V_{(0)}(\phi_{0})+\epsilon \left[ V'_{(0)}(u)\cdot \dfrac{\phi_{(1)}}{\phi '_{(0)}} +V_{(1)}(\phi_{(0)}) \right]+O(\epsilon ^{2}).
\end{equation}

\begin{figure}[H]
\includegraphics[scale=0.5]{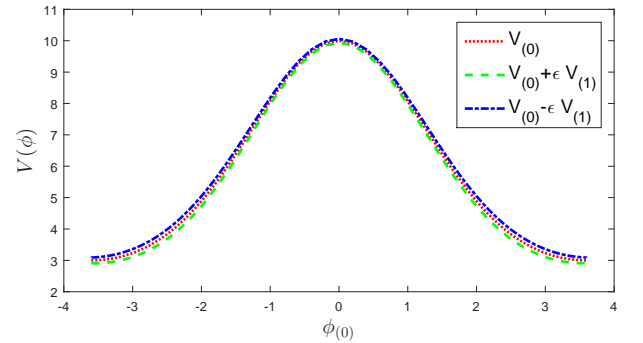}
\caption{Plot of the potential $V(\phi)$ and its first order correction, for the parameter values $\epsilon =0$.$01$, $c=0$.$5$ and $s=1$. }
\label{Fig:pertepspotencialc05}
\end{figure}

The potential $V(\phi)$ is plotted in the Fig. \ref{Fig:pertepspotencialc05}.

We obtain a $r^{2}=0.99975$ for $V_{0}$ and $V_{0}+\epsilon V_{1}$ and a $r^{2}=0.99975$, for $V_{0}$ and $V_{0}-\epsilon V_{1}$. \\ \\

The perturbation of the GEoS in the parameter $A$ around the vañue $-1/3$ leads also to bouncing solutions whose behavior in the scale factor and in the energy density are quite similar than the exact analytical solution found in this work. This also applied for the scenario in which a scalar field describe the matter content of the universe.

\section{Conclusions}

The study of GEoS has been very important in the exploration of new
scenarios in the very early phases of the universe like inflation,
or bouncing universe theories in which there are no initial singularities.

In this work we have found and exact analytical bouncing solution
for a closed universe filled with one fluid which obeys a GEoS of
the form $p= -1/3 \rho +B\rho^{1/2}$, where $B<0$ is a free
parameter. We have chosen the initial conditions that allow no
violation of NEC, which leads the parameter
$\omega$ to evolve with the cosmic time in the domain of
quintessence. For $t\rightarrow \pm \infty$, $\omega \rightarrow -1$ the
fluid behaves like a cosmological constant. We have also shown that
the well known de Sitter solution with positive curvature is
obtained as a particular case of our exact analytical bouncing
solution.

Another interesting feature of our result is the possibi-lity to interpret
the fluid ruled by the GEoS of Eq.(\ref{EoS1}) in terms of known fluids. In fact,
expanding the expressions for the pressure, $p$, and the
energy density, $\rho$, in terms of the scale factor and inserting them
into the GEoS, we obtained that the matter content can be
seen as the sum of the contributions coming from three fluids: a
cosmological constant, quintessence ($\omega = -2/3$) and the corresponding fluid
which arises from the particular value $\omega = -1/3 $.

Since the investigated fluid behaves effectively like quintessence, it is
possible to reinterpret the matter source in terms of an ordinary
scalar field $\phi$, minimally coupled to gravity with a positive
kinetic term and a potential $V(\phi)$.

We have solved for $\phi$ and $V(\phi)$ the set of coupled equations
(\ref{eccamppot}) by using Maclaurin expansions up to 14th order and, in paralell as an accuracy test,
we have computed them numerically from the implicit Eqs.(\ref{dphidu}) and (\ref{Vu}). A remarkable agreement was found by comparing the scalar field $\phi$ and the potential $V(\phi)$ obtained by both methods, whose precision is characterized by coefficients of determination $ r^{2}= 0.99997$ and $ r^{2}= 1$ respectively, (see Figs. (\ref{fig:discrevsseriephi0c05}) and (\ref{fig:discrevsserieV0c05})), which
holds in the common range.

Performing a high accuracy fit we have found an
analytical expression for the field and its scalar potential with
coefficients of determination typically of the order of one up to
$10^{-4}$ and $10^{-5}$ respectively (see Eqs. (\ref{fitfieldtanh}) and (\ref{Ajustegausscut}).
The shape of this potential can be very precisely
described by a Gaussian-type of function that has a bounded
domain given by the condition $ \phi_{min} < \phi < \phi_{max}$. With this exact analytical expression for the scalar potential we evaluated numerically the scale factor, founding a curve very close to the exact analytical bouncing solution already found. This represents a very astringent test on the accuracy of the numerical method used to compute the scalar potential $V(\phi)$.

We have also studied the structural stability of the analytical bouncing solution when
the GEoS is modified by including a perturbative term in the
standard linear coefficient $A$ (see Eq.(\ref{ecestadogen})). For sufficiently
small $\epsilon$, the scenario predicted by the analytic
solution is still preserved as the scale factor $a$ and the density
$\rho$ behave quite similar to the unperturbed solutions (see Eqs.
(\ref{relata_0a_1}) and (\ref{relatrho0rho1})). The shape of the scalar potential -introduced as a
possible source to generate the effective GeoS- also confirms the
unperturbed scenario: the absence of a spontaneous symmetry breaking
minimum, for $\epsilon$ small enough within the validity range of
the first order approximation.

In summary, the exact analytical bouncing
solution can be extended to a vicinity of $A=-1/3$, confirming a bouncing scenario beyond the particular value required for the exact solution. Moreover,
an analytical quintessence potential has been found by using a high
accuracy fit to the numerical data. A scalar field theory minimally
coupled to gravity and ruled by this potential leads to bouncing
solutions for closed universes, which does not present spontaneous
symmetry breaking.

The analytic scalar potential found in this article can further be used to study other interesting issues associated, like the consequences of considering perturbations in the background metric in the trivial minimum and absence of spontaneous symmetry breaking .

\section*{Acknowledgements}
\hspace{-10.5pt}This work was supported by CONICYT through Grant FONDECYT N$^0$
1140238 (NC). GP acknowledges partial funding by Dicyt-Usach Grant
N$^0$ 041531PA.

\vspace{0.5cm}
\setcounter{equation}{0}
\renewcommand{\theequation}{A\arabic{equation}}

\begin{center}
{\bf Appendix A: Maclaurin coefficients of $\phi$ and $V(\phi)$}
\end{center}

\hspace{-10.5pt}We will find the coefficients of Maclaurin series of the functions $\phi (u)$ and $V(\phi )$ and we will compute their convergence radius. These formal expansions up to order 14th are required in Figs. \ref{fig:discrevsseriephi0c05} and \ref{fig:discrevsserieV0c05}. 
We will firstly find the Maclaurin series of $\phi (u)$, and then we will invert this series to obtain the expansion for $u(\phi)$. These expressions will allow us to obtain the Maclaurin series of $V(\phi )$. In fact, starting with the definitions of the coefficients $\alpha _{n}$, $\beta _{n}$, and $c_{n}$ through the formal expansions

\begin{equation} 	
\begin{array}{l}
\phi (u)=\sum _{n=0}^{\infty} \alpha_{n} u^{n}, \quad
V (u)=\sum _{n=0}^{\infty} c_{n} u^{n},\vspace{2mm} \\
V (\phi )=\sum _{n=0}^{\infty} \beta_{n} \phi^{n}.\vspace{2mm} \\
\end{array}
\label{reptaylor}
\end{equation}
we will proceed finding the $\beta _{n}$ coefficients in three steps:

\begin{itemize}
\item First, we find the coefficients $\alpha _{n}$ of the Maclaurin series of $\phi (u)$, using its exact analytical expression.
\item Second, we find the coefficients $c _{n}$ of the Maclaurin series of $V (u)$ by using Eq.(\ref{Vu}).
\item third,  we find the coefficients $\beta _{n}$ of the Maclaurin series of $V(\phi )$ solving the implicit expression:

\begin{equation}
V(\phi )=\sum _{n=0}^{\infty } c _{n}u^{n}(\phi)
\label{relseries}
\end{equation}
\end{itemize}  
where $u(\phi)$ should be computed by inverting the expansion for $\phi(u)$ (see Eq. (\ref{reptaylor})). We begin by finding $\alpha _{n}$.  To find the Maclaurin series of $\phi (u)$ we integrate Eq.(\ref{dphidu}), which gives

\begin{equation}
\phi (u)=\pm \sqrt{2c}\int {\sqrt{\cosh (u)}\over \cosh (u)-c}du +s_{1} ,
\label{phiuintbounc} 
\end{equation}
with $s_{1}$ being an integration constant. Now we need the Maclaurin series of the integrand ${\sqrt{\cosh (u)}\over \cosh (u)-c}$, which can be derived by using the identity

\begin{equation}
f^{(n)}(u) = \sum _{k=1}^{n}{U_{k}(u)\over k!}F^{(k)}(y), \quad n\geq 1,
\label{IdenCompderivada} 
\end{equation}
where: \\ 
$U_{k}(u)=\sum _{i=1}^{k}\dfrac{(-1)^{i+1}}{(i-1)!}y^{i-1}(u)\left[\left( y^{k+1-i}(u)\right)^{(n)}(u)\right]  \times $\, \\   $\left. \quad \quad \quad \ \right. \left[\Pi _{j=1}^{i-1}(k+1-j)\right]$, and $f(u)=(F\circ y)(u)$. \\ 

Note that each $U_{k}$ also depends of $n$. If we consider $F(y)={\sqrt{y}\over y-c}$ and $\psi (u)=\cosh (u)$, we obtain that $f(u)$ which is precisely the integrand of Eq.(\ref{phiuintbounc}).\\ 

Now, reemplacing $y=\cosh(u)$ into Eq.(\ref{IdenCompderivada}) and evaluating $u$ at $u=0$, we obtain 

\begin{equation}
f^{(n)}(0)=\sum _{k=1}^{n}{U_{k}(0)\over k!}F^{(k)}(\cosh (0)), 
\label{fderivadabounc} 
\end{equation}
where: \, \\
$U_{k}(0) =\sum _{i=1}^{k}{(-1)^{i+1}\over (i-1)!}     \left[\left( \cosh ^{k+1-i}(u)\right)^{(n)}(0)\right]\times$ \, \\ $\left. \quad \quad \quad \right. \left[\prod _{j=1}^{i-1}(k-j+1)\right]$. \\ Using Leibniz's rule for the derivative of a product

\begin{equation}
(AB)^{(k)}(y)=\sum _{i=0}^{k}{k \choose i}A^{(k-i)}(y)B^{(i)}(y),
\label{Fderivadabounc} 
\end{equation}
with $A(y)=\sqrt{y}$ and $B(y)=(y-c)^{-1}$, and the expressions

\begin{equation}
\begin{array}{l}
A^{(n)}(y)={(-1)^{n+1}(2n-3)!!\over 2^{n}}y^{{1\over 2}-n}, \vspace{2mm} \\
B^{(n)}(y)=n!(-1)^{n}(y-c)^{-n-1} .
\end{array}
\label{AyBdeFbounc} 
\end{equation}

we obtain

\begin{equation}
F^{(k)}(\cosh (0))=\sum _{i=0}^{k}{k!\over (k-i)!}\cdot {(2k-2i-3)!!\over 2^{k-i}}{(-1)^{k+1}\over (1-c)^{i+1}}.
\label{Fderivadabounc3} 
\end{equation}

In order to obtain the functions $U_{k}(0)$ we use the identities

\begin{equation}
\begin{array}{l}
\cosh ^{2k}(u)=2^{-2k}\left[ \sum _{i=0}^{k-1}2{2k \choose i}\cosh (2(k-i)u) + {2k \choose k} \right], \vspace{2mm} \\
\cosh ^{2k-1}(u)=2^{2-2k}\sum _{i=0}^{k-1}{2k-1 \choose i}\cosh ((2k-2i-1)u).
\end{array}
\label{Idendercoshkesima} 
\end{equation}

From which we obtain for $l\geq 1$ and $n\geq 1$

\begin{equation}
\begin{array}{l}
 \left(\cosh ^{l}(u)\right)^{(2n-1)} (0)=   0,\vspace{2mm} \\
\left( \cosh ^{2l}(u)\right)^{(2n)}(0)=2^{-2l}\sum _{i=0}^{l-1}2{2l \choose i}(2(l-i))^{2n},\vspace{2mm} \\
\left(\cosh ^{2l-1}(u)\right)^{(2n)} (0)=2^{2-2l}\sum _{i=0}^{l-1}{2l-1\choose i}(2l-2i-1)^{2n}.
\end{array}
\label{dernesparfbounc} 
\end{equation}

Moreover, using Eq.(\ref{Fderivadabounc3}), Eq.(\ref{dernesparfbounc}) and Eq.(\ref{fderivadabounc}), we find

\begin{equation}
\begin{array}{ll}
f^{(m)}(0)&= f^{(2n)}(0)=\sum _{k=1}^{2n}(-1)^{k+1}U_{k}(0)\delta _{k}\vspace{2mm} \\
&=\sum _{k=1}^{n}U_{2k-1}(0)\delta _{2k-1}-\sum _{k=1}^{n}U_{2k}(0)\delta _{2k},\\ \\
\end{array}
\label{fderivadabounc2} 
\end{equation}
where  \quad $\delta_{k}=\sum _{i=0}^{k}{(2k-2i-3)!!\over (k-i)!2^{k-i}}(1-c)^{-i-1}$, which leads to the expression

\begin{equation}
f^{(2n)}(0)=\sum _{k=1}^{n}(U_{2k-1}(0)\delta _{2k-1}-U_{2k}(0)\delta _{2k}),
\label{fderivadabounc3} 
\end{equation}
where $\ n\geq 1$, and \\
$U_{2k}(0)=\sum _{i=1}^{k}\left\{ {\eta_{i,k}\over (2i-2)!}\left[ \prod _{j=1}^{2i-2}(2k-j+1)\right]\right.$\, \\ 
$\left. \quad \quad \quad \quad -{\gamma _{i,k}\over (2i-1)!}\left[ \prod _{j=1}^{2i-1}(2k-j+1)\right]\right\}$, \, \\
$U_{2k-1}(0)=(2k-1)+\sum _{i=1}^{k-1}\left\{ {\gamma _{i,k}\over (2i-2)!}\left[ \prod _{j=1}^{2i-2}(2k-j)\right]\right.$\, \\
 $\left. \quad \quad \quad \quad \quad -{\eta _{i+1,k}\over (2i-1)!}\left[ \prod _{j=1}^{2i-1}(2k-j)\right]\right\}$,\, \\
$\eta _{i,k}=2^{2i-2k-1}\sum _{j=0}^{k-i}{2(k+1-i)\choose j}(2(k+1-i-j))^{2n}$, \, \\
$\gamma _{i,k}=2^{2i-2k}\sum _{j=0}^{k-i}{2k+1-2i\choose j}(2k+1-2i-2j)^{2n}$. \\

Finally, using the above results for $f(u)$ we obtain 

\begin{equation}
f(u)={\sqrt{\cosh (u)}\over \cosh (u)-c}={1\over 1-c}+\sum _{n=1}^{\infty }{f^{(2n)}(0)\over (2n)!}u^{2n}.
\label{fubounc} 
\end{equation}

Now, inserting Eq.(\ref{fubounc}) into Eq.(\ref{phiuintbounc}) it follows

\begin{equation}
\begin{array}{l}
\phi (u)=\pm \sqrt{2c}\left[ {u\over 1-c}+\sum _{n=1}^{\infty }{f^{(2n)}(0)\over (2n)!} {u^{2n+1}\over 2n+1}\right]+s_{1}.\\
\end{array}
\label{phiuserie} 
\end{equation}

From the above expression, one can compute the first coefficients of the Maclaurin series of $\phi (u)$ 

\begin{equation}
\begin{array}{l}
\alpha _{0}= s_{1}, \quad
\alpha _{1}=\pm \frac{\sqrt{2c}}{1-c}, \quad
\alpha _{2}=0, \vspace{2mm} \\
\alpha _{3}=\mp \dfrac{\sqrt{2c}}{12}\dfrac{1+c}{(1-c)^{2}},\quad
\alpha_{4}=0.
\end{array}
\label{coefalphasubkparbounc} 
\end{equation}

Using the fact that the convergence radius of a power series  of an analytic function is the distance from the center of the power series to closest singularity and that $0<c<1$, we conclude that the convergence radius of Maclaurin series of $\phi(u)$ of Eq.(\ref{phiuserie}) is equal to $\arccos (c)$. \\

Now, we will compute the coefficients $\beta _{n}$ of the Maclaurin series of $V(\phi )$. To this aim, we use the coefficients $c_{n}$, which can be obtained in a similar way to how the coefficients $\alpha _{n}$ were computed since we can rewrite Eq.(\ref{Vu}) as

\begin{equation}
\begin{array}{ll}
V(u)=&{9B^{2}\over 4}\left\{ 1\right.\vspace{2mm} \\
& \left.+{5c\over 3}\left[ \left( 1+{2c\over 5}\left( {1\over \cosh (u)-c}\right) \right) \left( {1\over \cosh (u)-c}\right)\right]\right\}.
\end{array}
\end{equation}

A straightforward computation yields

\begin{equation} 
V(u)={9B^{2}\over 4}\left[ {-c+3\over 3(1-c)^{2}}+{5\over 3}c\sum _{n=1}^{\infty }{Y^{(2n)}\over (2n)!}u^{2n}\right], \vspace{2mm} \\
\label{Vuserie} 
\end{equation}
where: \\ 
$Y^{(2n)}= {2c\over 5}\sum _{k=1}^{n-1}{2n\choose 2k}g_{2(n-k)}g_{2k} + \left( 1+{4c\over 5(1-c)}\right) g_{2n}$, with $Y^{(2n-1)}=0$ for $n\geq 1$, and constants $g_{2n}$ given by

\begin{equation}
g_{2n}=\sum _{k=1}^{n}(U_{2k}\delta_{2k}-U_{2k-1}\delta _{2k-1}), \quad n\geq 1,
\label{derenesgbounc3} 
\end{equation}
with\\
$U_{2k} = \sum _{i=1}^{k}\left\{ {\eta_{i,k}\over (2i-2)!}\left[ \prod _{j=1}^{2i-2}(2k-j+1)\right]\right.$ \\    $\left. \quad \quad \quad  -{\gamma_{i,k}\over (2i-1)!}\left[ \prod _{j=1}^{2i-1}(2k-j+1)\right]\right\}$, \\
 $U_{2k-1}=(2k-1)+\sum _{i=1}^{k-1}\left\{ {\gamma_{i,k}\over (2i-2)!}\left[ \prod _{j=1}^{2i-2}(2k-j)\right]\right.$, \\ 
$ \left. \quad \quad \quad \quad -{\eta_{i+1,k}\over (2i-1)!}\left[ \prod _{j=1}^{2i-1}(2k-j)\right]\right\}$, \\
$\eta _{i,k}=2^{2i-2k-1}\sum _{j=0}^{k-i}{2(k+1-i)\choose j}(2(k+1-i-j))^{2n}$, $\gamma _{i,k}=2^{2i-2k}\sum _{j=0}^{k-i}{2k+1-2i\choose j}(2k+1-2i-2j)^{2n}$, \\
$\delta _{k}=(1-c)^{-1-k}$. \\

The expansion of Eq.(\ref{Vuserie}) has a convergence radius $r_{c}=\arccos(c)$.
Now we are equipped with the required relations to finally obtain the coefficients $\beta _{n}$ of the expansion of $V(\phi )$. We use Eq.(\ref{relseries}) together with the following identity  

\begin{equation}
\left( \sum _{k=0}^{\infty }a_{k}u^{k}\right)^{n}=\sum _{k=0}^{\infty }c_{(k,n)}u^{k}, \quad n\geq 0, 
\label{Idenpotenciaserie} 
\end{equation}
where $c_{(0,n)}=a_{0}^{n}$, and $c_{(m,n)}={1\over ma_{0}}\sum _{k=1}^{m}(kn-m+k)a_{k}c_{(m-k,n)}$ for $m\geq 1$. For the particular case $s_{1}=0$, which leads to a well defined and unique solution for the parameters $\beta _{n}$, we obtain 

\begin{equation}
\begin{array}{c}
c_{0}=\beta _{0}, \quad
c_{n}=\sum _{k=1}^{n}\beta _{k}c_{(n-k,k)} \ \text{for } \ n\geq 1.
\end{array}
\label{sistrealresolbounc} 
\end{equation}

To obtain the coefficient $\beta _{k}$, we used that $\phi (u)$ is an odd function while $V(u)$ is an even function of the argument. Thus solving Eq.(\ref{sistrealresolbounc}) for $\beta _{k}$ we obtain

\begin{equation}
\begin{array}{l}
V(\phi )=\sum_{k=0}^{\infty}\beta _{k}\phi ^{k},\quad  \beta _{0} =c_{0}, \quad \beta _{2n-1}=0, \vspace{2mm} \\
\beta _{2n}={c_{2n}\over c_{(0,2n)}}-{\sum _{i=1}^{n-1}\beta _{2i}c_{(2n-2i,2i)}\over c_{(0,2n)}}, \quad
n\geq 1.
\end{array}
\label{coefbetasubkgenbounc} 
\end{equation}

The convergence radius of $V (\phi )$ is $r_{c}=\phi (\arccos (c))$. Finally, we evaluate explicitly the first $\beta _{n}$-coefficients from the relations of Eq.(\ref{coefbetasubkgenbounc})

\begin{equation}
\begin{array}{l}
\beta _{0}=  \dfrac{3B^{2}(3-c)}{4(1-c)^{2}},  \quad
\beta _{1}=0, \quad
\beta_{2}=-  \dfrac{3B^{2}(5-c)}{16(1-c)}, \vspace{2mm} \\
\beta_{3}=0, \quad
\beta _{4}=\dfrac{B^{2}(15+4c+c^{2})}{128c}.
\end{array}
\label{3primbetasubkparbounc} 
\end{equation}

\setcounter{equation}{0}
\renewcommand{\theequation}{B\arabic{equation}}

\begin{center}
{\bf Appendix B: Differential equations for $a$ and $\rho$ up to first order }
\end{center}
 
\vspace{0.5cm}
\hspace{-10.5pt}We will compute the coefficients $A$, $B$, $C$, $D$, $E$ and $F$ that appear in Eq.(\ref{Sistema}). We first compute the function $p_{(1)}$ appearing in Eqs.(\ref{ecestadogen}) and(\ref{sistgen}):

\begin{equation}
p_{(1)}=\rho _{(0)}-\dfrac{\rho _{(1)}}{3}-\dfrac{\rho _{(1)}}{s\sqrt{3}\rho _{(0)}^{1/2}},
\label{ecpertparaP}
\end{equation}
which corresponds to Eq.(\ref{solfinaleps1ord}). Now, in order to obtain the coefficients A, B, C, D, E and F, we insert Eqs.(\ref{sistgen}) and (\ref{solord0}) into Eq.(\ref{constraint}), which leads to the following expression

\begin{equation}
a_{(1)}'=s a_{(1)}\left( \dfrac{a_{(0)}'}{sa_{(0)}}+\dfrac{s}{a_{(0)}'a_{(0)}}\right) + \dfrac{s^{2}\rho _{(1)}a_{(0)}^{2}}{6a_{(0)}'},
\label{ecpertparaa2}
\end{equation}
where the ``prime'' ({ \large $_{'} \ $}) denotes the derivative with respect to $u$. Now, inserting $a_{0}(u)$, $\rho _{0}(u)$ and $a'_{0}(u)$ in the above equation we obtain

\begin{equation}
a'_{1}(u)=A(u)a_{1}(u)+B(u)\rho _{1}(u),
\end{equation}
where

\begin{equation}
\begin{array}{l}
A(u) = \dfrac{\cosh ^{2}(u)}{\sinh (u)(\cosh (u)-c)}, B(u)=\dfrac{s^{3}(\cosh (u)-c)^{2}}{6\sinh (u)}.
\end{array}
\label{coeffAandB}
\end{equation}

Moreover, if we use Eqs.(\ref{sistgen}) and  (\ref{ecconsgen}), we obtain

\begin{equation}
\begin{array}{ll}
 \rho _{(1)}'&=\rho _{(1)} \dfrac{3a_{(0)}'}{a_{(0)}}\left( \dfrac{1}{s\sqrt{3}\rho _{(0)}^{1/2}}-\dfrac{2}{3}\right)-\dfrac{3a_{(1)}'}{a_{(0)}}\left( \rho _{(0)}+p_{(0)}\right) \vspace{2mm} \\ 
 &\quad +\dfrac{3a_{(1)}a_{(0)}'}{a_{(0)}^{2}}\left( \rho _{(0)}+p_{(0)}\right) -\dfrac{3a_{(0)}'}{a_{(0)}}\rho _{(0)}.
\end{array}
\label{rho_1nocomple}
\end{equation}

When the functions $a_{0}(u)$, $\rho _{0}(u)$, $p_{0}(u)$ and $a'_{0}(u)$ are inserted into Eq.(\ref{rho_1nocomple}), $\rho '_{1}(u)$ can explicitly be  expressed as

\begin{equation}
\rho' _{1}(u)=C(u)a'_{1}(u)+D(u)a_{1}(u)+E(u)\rho _{1}(u)+F(u)
\end{equation}
where

\begin{equation*}
C(u)=\dfrac{-6c\cosh (u)}{s^{3}(\cosh (u)-c)^{3}},
\end{equation*}
\begin{equation*} 
D(u)=\dfrac{6c\cosh  (u)\sinh (u)}{s^{3}(\cosh (u)-c)^{4}},
\end{equation*}
\begin{equation*}
E(u)= -\dfrac{\sinh (u)(\cosh (u)+c)}{\cosh (u)(\cosh (u)-c)}, 
\end{equation*}
\begin{equation}
F(u)=-\dfrac{9\cosh ^{2}(u)\sinh (u)}{s^{2}(\cosh (u)-c)^{3}}.
\label{coeffC,D,EandF}
\end{equation}

\end{document}